\documentclass[a4paper]{spie}

\usepackage{aas_macros}
\usepackage{amsmath,amsfonts,amssymb}
\usepackage{graphicx}
\usepackage[colorlinks=true, allcolors=blue]{hyperref}

\usepackage[]{subfigure}

\title{Calibration of the IXPE instrument}

\author[a]{Fabio Muleri}
\author[a]{Carlo Lefevre}
\author[a]{Raffaele Piazzolla}
\author[a]{Alfredo Morbidini}
\author[a]{Fabrizio Amici}
\author[a]{Primo Attin\'{a}}
\author[d]{Mauro Centrone}
\author[a]{Ettore Del Monte}
\author[a]{Sergio Di Cosimo}
\author[a]{Giuseppe Di Persio}
\author[a]{Yuri Evangelista}
\author[a]{Sergio Fabiani}
\author[a]{Riccardo Ferrazzoli}
\author[a]{Pasqualino Loffredo}
\author[c]{Luca Maiolo}
\author[c]{Francesco Maita}
\author[a]{Leandra Primicino}
\author[a]{John Rankin}
\author[a]{Alda Rubini}
\author[a]{Francesco Santoli}
\author[a]{Paolo Soffitta}
\author[a]{Antonino Tobia}
\author[a]{Alessia Tortosa}
\author[b]{Alessio Trois}
\author[ ]{on behalf of the IXPE team}

\affil[a]{INAF-IAPS, Via del Fosso del Cavaliere 100, I-00133 Roma, Italy}
\affil[b]{INAF-OAC Via della Scienza 5 - 09047 Selargius (Cagliari), Italy}
\affil[c]{CNR-IMM via del Fosso del Cavaliere 100, I-00133, Roma, Italy}
\affil[d]{INAF-OAR, Via di Frascati, 33, I-00040 Monte Porzio Catone, Italy}

\authorinfo{Further author information: (Send correspondence to Fabio Muleri)
\\Fabio Muleri: E-mail: fabio.muleri@iaps.inaf.it, Telephone: +39-0649934565}

\begin{document} 
\maketitle

\begin{abstract}
IXPE scientific payload comprises of three telescopes, each composed of a mirror and a 
photoelectric polarimeter based on the Gas Pixel Detector design. The three focal plane detectors, 
together with the unit which interfaces them to the spacecraft, are named IXPE \emph{Instrument}
and they will be built and calibrated in Italy; in this proceeding, we will present how IXPE 
Instrument will be  calibrated, both on-ground and in-flight. The Instrument Calibration Equipment 
is being  finalized at INAF-IAPS in Rome (Italy) to produce both polarized and unpolarized 
radiation, with a precise knowledge of direction, position, energy and polarization state of the 
incident beam. In flight, a set of four calibration sources based on radioactive material and 
mounted on a filter and calibration wheel will allow for the periodic calibration of all of the 
three IXPE focal plane detectors independently. A highly polarized source and an unpolarized one  
will be used to monitor the response to polarization; the remaining two will be used to calibrate 
the gain through the entire lifetime of the mission.
\end{abstract}


\section{Introduction} \label{sec:Introduction}

The Imaging X-ray Polarimetry Explorer (IXPE, see O'Dell et al. in this same volume\cite{Odell2018}) 
will be launched in April 2021 carrying on-board the first instrument dedicated to X-ray 
polarimetry in decades. The mission, selected in the context of the NASA Astrophysics Small 
Explorer (SMEX) program in January 2017, is a collaboration with the Italian Space Agency (ASI) 
which will provide with INAF and INFN the focal plane instrumentation. This comprises three flight 
\emph{Detection Units} (DUs), plus a spare, and a \emph{Detector Service Unit} (DSU). The former, 
built by INFN-Pisa, is the unit which contains the device sensitive to X-ray polarization, which is 
based on the photoelectric effect and on the Gas Pixel Detector design developed in Italy for nearly 
20~years \cite{Costa2001,Bellazzini2006,Bellazzini2007}. The DU also includes a \emph{Filter and 
Calibration Wheel} (FCW), which hosts filters for special observations and four calibration sources 
to monitor GPD performance in flight, as described in Section~\ref{sec:FCW}. The DSU contains the 
electronics required to manage the DUs and interface them to the spacecraft.

DUs will be calibrated and functionally tested in Italy, at INAF-IAPS, before delivery to NASA 
Marshall Space Flight Center (NASA-MSFC), where they will be calibrated together with the 
mirror built by MSFC for the telescope calibration. While the latter will be dedicated to an 
end-to-end calibration of the IXPE instrumentation before integration on the spacecraft, DU 
calibration will feature an extensively characterization of the scientific performance of the 
GPD. The equipment which will used at this aim are collectively named \emph{Instrument 
Calibration Equipment} (ICE) and they will be described in Section~\ref{sec:ICE}.

\section{Instrument on-ground calibrations}

Instrument calibration will be devoted to:
\begin{itemize}
\item  measure the modulation factor as a function of energy to better than 1\% of its value; the 
same measurements will allow also deriving how the energy resolution varies with energy. 
Measurements will be carried out with a beam of collimated, polarized and monochromatic photons at 
different energies. The spot will be about 0.4 mm, that is, comparable to the spot from a real X-ray 
telescope. For a few reference energies the measurements will be repeated on a grid of positions to 
map all the sensitive area of the GPD, repeating periodically a measurement in a reference position 
for checking possible intrinsic variations of the beam. The procedure is already in use at 
INAF-IAPS for the calibration of GPD prototypes.
\item measure the spatial resolution as a function of energy. Measurements will be carried out 
with a collimated, pencil beam of tens $\mu$m size, that is much smaller than the GPD spatial 
resolution, on a grid of positions and at a few energies. Also in this case, the procedure was 
already used with the GPD\cite{Soffitta2013}.
\item check the absence of a significant spurious polarization to $\lesssim$0.1\%. Measurements 
will be repeated at a few energies in several positions of the sensitive area.
\item check the relation between the expected and measured angle of polarization. Measurements 
with polarized radiation at a few energies will be repeated in the same point of the detector 
changing the angle of polarization to verify the relation between the expected and the measured 
value.
\item Map the gain. The gain of the Gas Electron Multiplier can change with the 
position of a factor $\pm$ few tens \%\cite{Tamagawa2009}. The gain will be mapped by fully 
illuminating the GPD with at least two sources at different energies.
\item Measure the efficiency. The efficiency of the detector will be measured by means of the 
comparison of the detected rate with a devices of known efficiency; the measurements will be 
carried out in one representative point of the detector and repeated at several different energies.
\end{itemize}

An exemplary list of measurements which will be carried out during Instrument calibration is listed 
in Table~\ref{tab:CalibrationPlan}. The total time for calibration of each DU is 45~days.

\begin{table}[htbp]
\begin{center}
\includegraphics[width=0.6\textwidth]{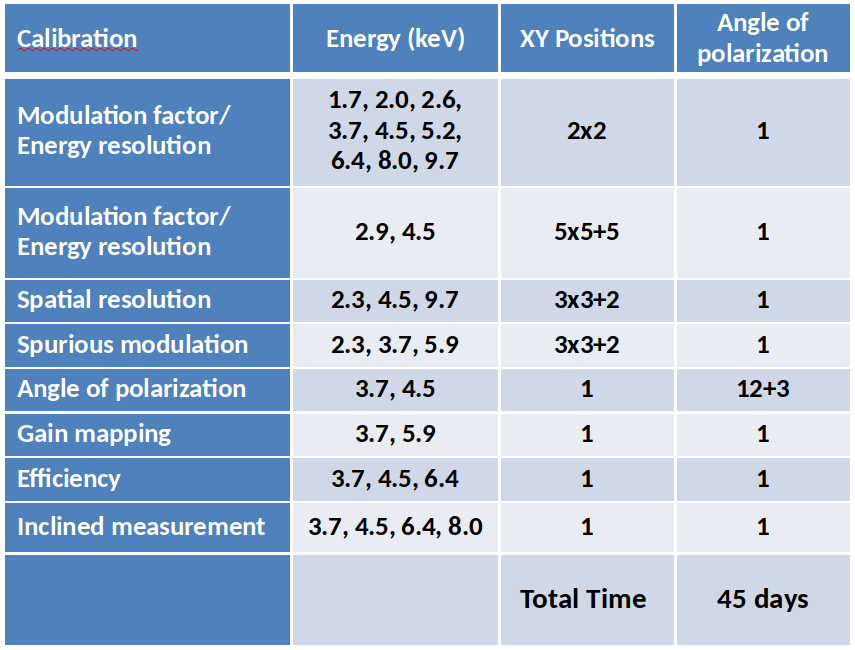}
\end{center}
\caption{Exemplary calibration plan for Instrument.} 
\label{tab:CalibrationPlan}
\end{table}

\subsection{The Instrument Calibration Equipment} \label{sec:ICE}

The Instrument Calibration Equipment (ICE) comprises the items which are used for Instrument 
calibration and functional tests. In particular, it includes:
\begin{itemize}
\item the X-ray sources used for illuminating the detector. Each source emits X-ray photons at 
known energy and with known polarization degree and angle. The direction of the beam, the direction 
of polarization for polarized sources, and its position can be measured with respect to the GPD 
inside the DU and aligned and moved as necessary. 
\item the test detectors which are used to characterized the beam before DU calibration and as a 
reference for specific measurements (e.g., the measurement of quantum efficiency).
\item all the electrical and mechanical equipment required to support the DU and the calibration 
sources, monitor the relevant diagnostic parameters and assure safe operations during calibrations. 
This includes also the clean environment (class better than 100,000) in which DU will be calibrated.
\end{itemize}

\begin{figure} [ht]
\begin{center}
\begin{tabular}{c} 
\includegraphics[width=0.75\textwidth]{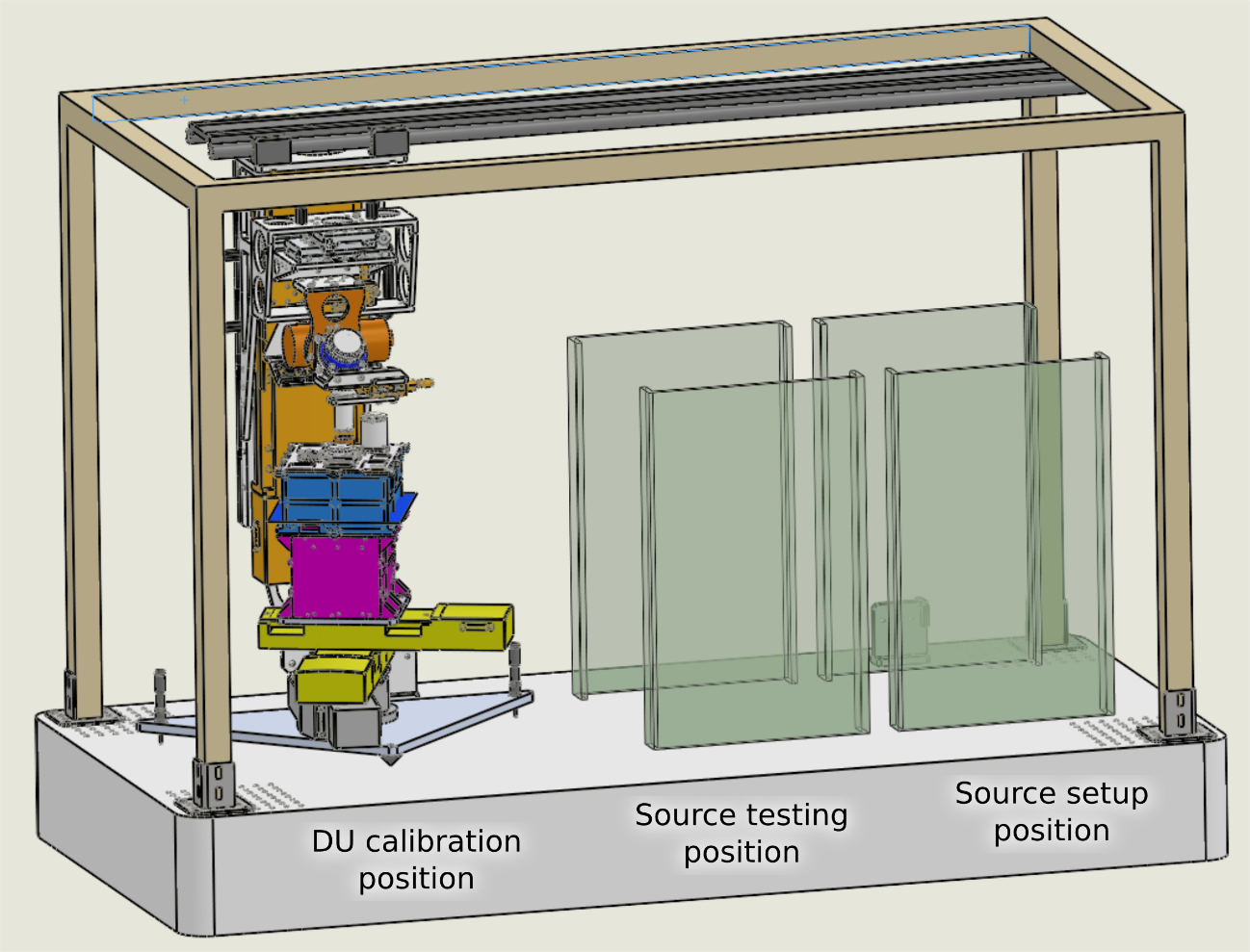}
\end{tabular}
\end{center}
\caption{Sketch of the calibration equipment for DU calibration.}
\label{fig:ICE}
\end{figure}

\begin{figure}[htbp]
\begin{center}
\subfigure[\label{fig:ICE_tower}]{\includegraphics[angle=0,totalheight=6.2cm]{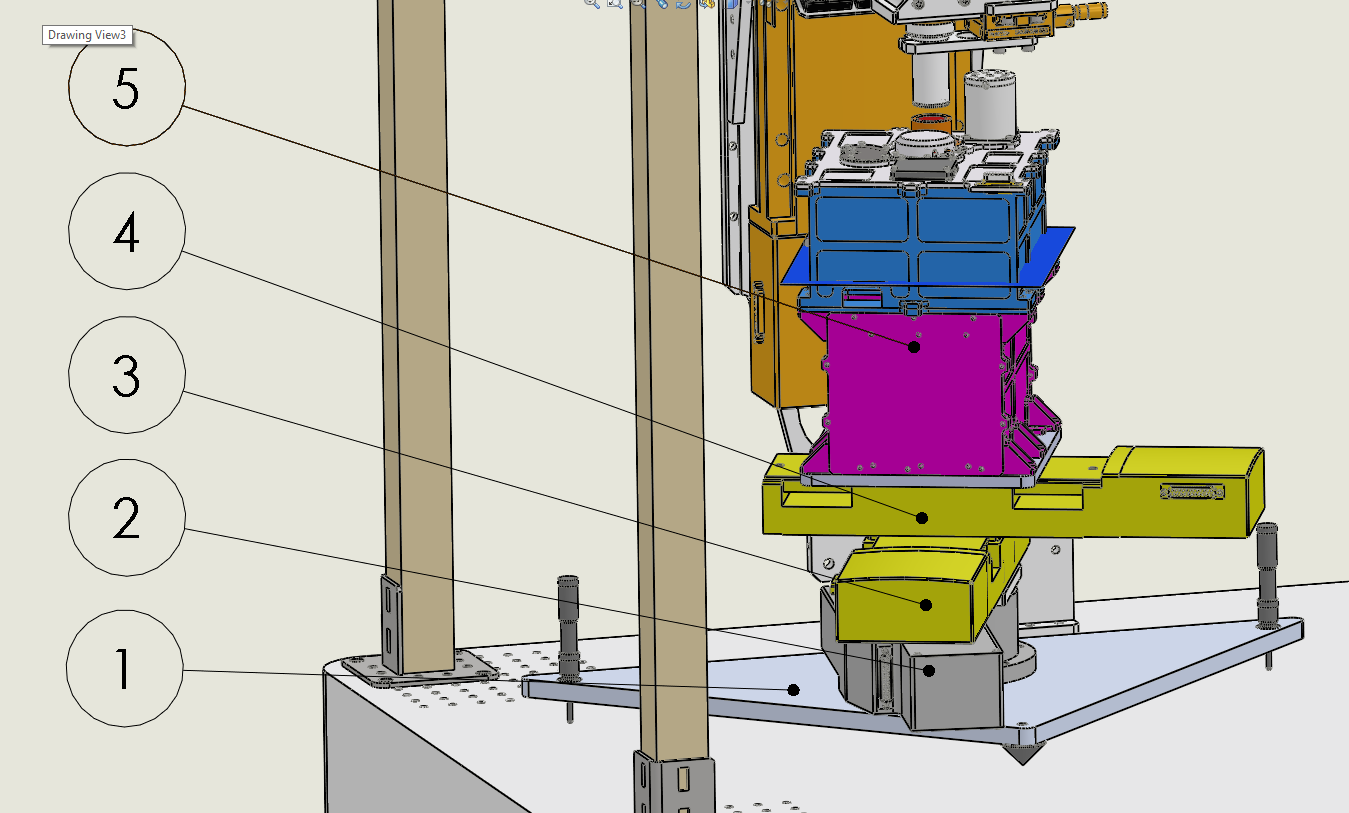}}
\hspace{3mm}
\subfigure[\label{fig:ICE_top}]{\includegraphics[angle=0,totalheight=6.2cm]{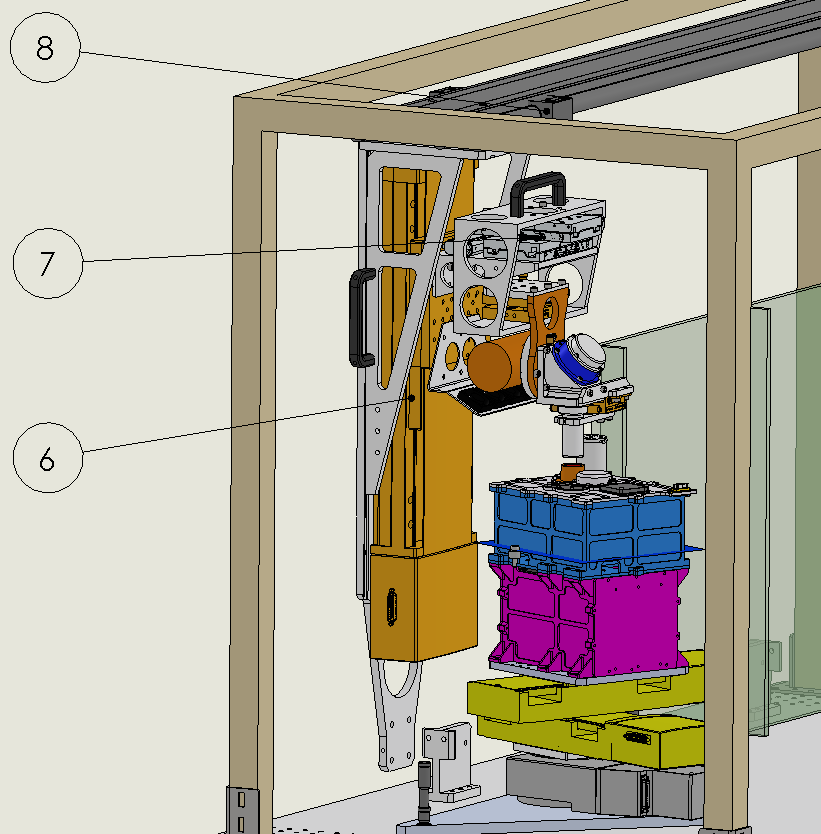}}
\end{center}
\caption{({\bf a}) Detailed view of the tower supporting the DU. ({\bf b}) Overview of the 
source assembly.}
\end{figure}

A drawing of the ICE when the polarized source is mounted is shown in Figure~\ref{fig:ICE}. The 
DU will be mounted in the ICE without the stray-light collimator and the UV filter, to limit the 
distance between the X-ray source and the GPD and hence air absorption and beam divergence. The DU 
is placed on the top of a tower (see Figure~\ref{fig:ICE_tower}) which allows to:
\begin{itemize}
\item move the DU (item 5 in the figure) on the plane orthogonal to the incident beam with an 
accuracy of $\pm2~\mu$m (over a range of 100~mm) to map the GPD sensitive surface. These two stages 
(items 4 and 3) are named \emph{xdu} and \emph{ydu}.
\item rotate the DU on the plane orthogonal to the incident beam with an accuracy of 
$\pm$7~arcsec, to test the response at different polarization angle values and to average residual 
polarization of unpolarized sources, if necessary. This stage (item 2) is named $\epsilon$.
\item tip/tilt align the orthogonal direction of the GPD to the incident beam (item~1). Two out of 
the three feet of the tip/tilt plate (called $\eta_1$ and $\eta_2$) will be manual micrometers, but 
one ($\eta_0$) will be motorized to carry out automatically measurements with the beam off-axis of a 
series of known angles, between $<$1~degree and about 5~degrees, e.g., to simulate the focusing of 
X-ray mirror shells.
\end{itemize}

The X-ray source is mounted on a mechanical support which allows to adjust its position and 
inclination with respect to the DU (see Figure~\ref{fig:ICE_top}). A manual translation stage 
($\chi$, item 8) allows to move the source assembly in three separate positions: (1) calibration 
with DU; (2) source characterization with test detectors; (3) source set-up and mounting (see 
Figure~\ref{fig:ICE}). The source assembly is fixed when in position with two brakes mounted at its 
top and bottom ends. A vertical motorized stage ($\nu$, item 6) with range 300~mm and accuracy 
$\pm5~\mu$m allows to move the source at different heights and minimize the distance of the source 
to the DU; software and hardware limitations will avoid the source to hit the DU even in case of 
wrong commanding. Two motorized linear stages (item~7) mounted in XY configuration on the 
plate of the vertical stage, named \emph{xso} and \emph{yso}, allows to move the X-ray source and 
center the beam on the axis of rotation of DU rotation stage $\epsilon$, to avoid moving the spot 
while rotating this stage. The uppermost of the XY stages is the mechanical interface for mounting 
polarized and unpolarized calibration sources on the source assembly.

\begin{figure}[ht]
\begin{center}
\begin{tabular}{c} 
\includegraphics[width=0.9\textwidth]{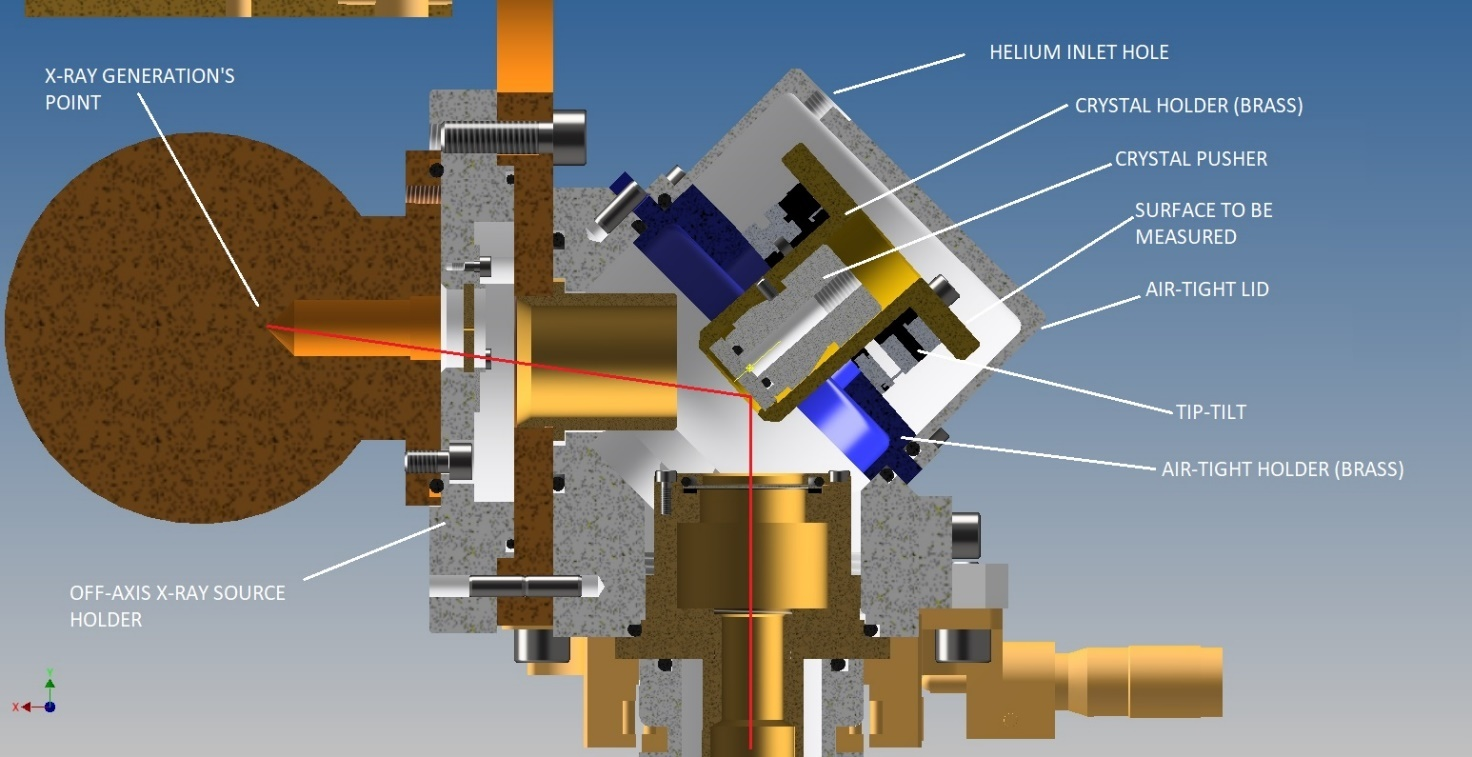}
\end{tabular}
\end{center}
\caption{Cross section of the polarized calibration source.} 
\label{fig:ICE_source}
\end{figure}

\begin{figure}[ht]
\begin{center}
\begin{tabular}{c} 
\includegraphics[width=0.8\textwidth]{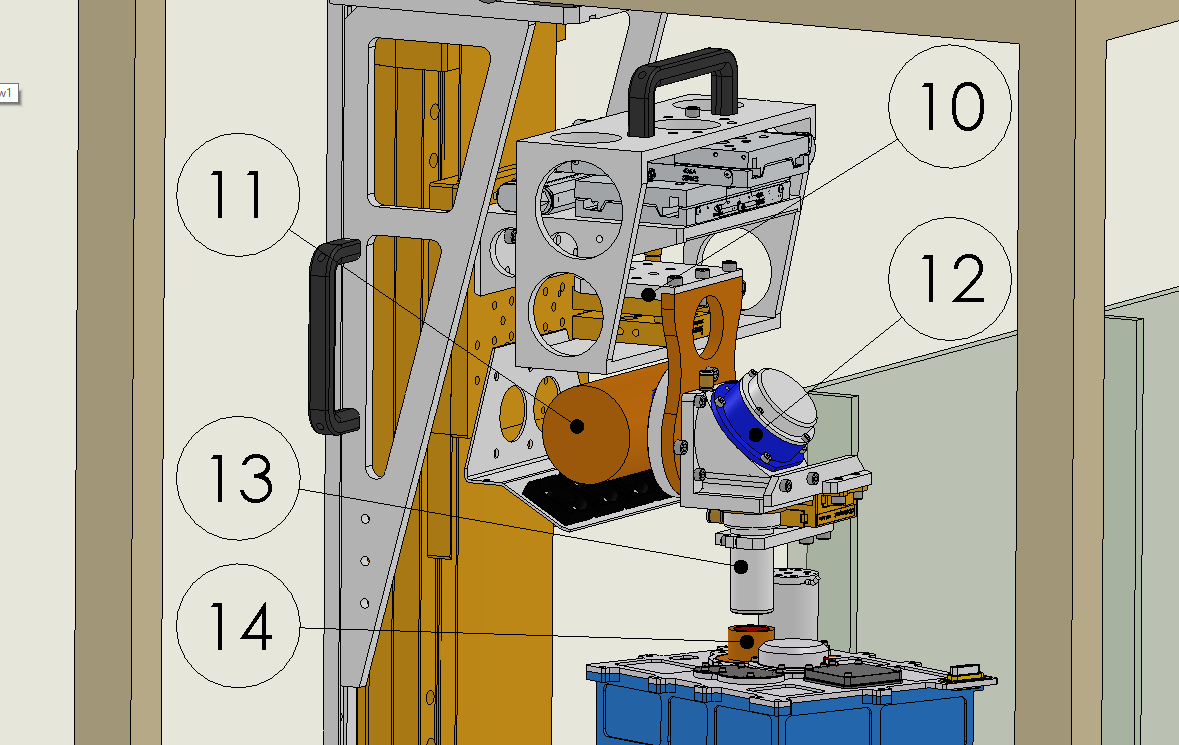}
\end{tabular}
\end{center}
\caption{ Detailed view of the polarized source assembly.} 
\label{fig:ICE_Helium}
\end{figure}

A cut-out view of the polarized source with Oxford Series 5000 X-ray tube mounted is reported in 
Figure~\ref{fig:ICE_source}. Polarized photons are produced by means of Bragg diffraction at nearly 
45 degrees on different crystals to cover the whole DU energy range, with a design based 
on the heritage of the calibration facility used at IAPS for GPD calibration for 10 
years\cite{Muleri2008b}. Crystal orientation, which sets the energy, the polarization degree and 
angle, and the direction of the beam, can be initially adjusted with a manual stage and, once fixed, 
measured with high accuracy. Crystal holder are designed so that the surface of the crystal is 
parallel by design to a surface accessible after source integration; the orientation of such a 
surface is measured with a measurement arm and referred to the GPD orientation by means of the 
alignment references in the DU. A diaphragm with aperture from 2 mm to 25~$\mu$m will be mounted at 
the end of the source extension which will avoid mechanical interference with DU parts extending 
from the lid. Diaphragm position will be manually adjustable to center the spot diffracted by the 
crystal, with a procedure based on the imaging of the X-ray beam with test detector and an alignment 
diaphragm. Inner part of the source will be made of brass to reduce scattering of impinging photons, 
and collimators at capillary plates can be mounted to constrain the direction of incident and 
diffracted radiation. A tip/tilt stage allows for beam direction adjustment in addition to the DU 
tip/tilt plate. Inner chamber of the source will be flown with helium to reduce the severe air 
absorption of X-rays in the IXPE energy range. Helium will be flown also inside a cylinder mounted 
on the lid of the DU (item 14 in Figure~\ref{fig:ICE_Helium}) and extending inside it, to reduce air 
absorption inside the DU. Helium will not be dispersed inside the DU to avoid discharges from the 
GPD Titanium top frame, which is at high voltage.

The polarized source can host also X-ray tubes different from Oxford Series 5000, e.g. Hamamatsu 
Head-on N7599 series, without disassembling the crystal and the collimator from the source 
assembly. Moreover, a source, based on the Compton scattering of X-ray photons in a Lithium rod 
encapsulated in Beryllium, will be available to generate highly-polarized continuum radiation which 
is representative of the spectrum of a typical astrophysical source. The list X-ray tubes, crystals 
and energies available for Instrument calibrations is shown in  Table~\ref{tab:Crystals}.

\begin{table}[htbp]
\begin{center}
\begin{tabular}{ccc} 
Energy (keV) & Crystal & X-ray tube \\
\hline
\hline
1.7 & Ammonium Dihydrogen Phosphate - ADP & Oxford 5000 Series, Titanium, 50 W \\
\hline
2.0 & Pentaerythritol - PET & Oxford 5000 Series, Titanium, 50 W \\
\hline
2.3 & Rhodium & Oxford 5000 Series, Molybdenum, 50 W \\
\hline
2.6 & Graphite & Oxford 5000 Series, Titanium, 50 W \\
\hline
2.7 & Germanium (111) & Oxford 5000 Series, Rhodium, 50 W \\
\hline
3.0 & Silicon (111) & Oxford 5000 Series, Silver, 50 W \\
\hline
3.7 & Aluminum & Head-on Hamamatsu, Calcium, 0.2 W \\
\hline
4.5 & Calcium fluorite & Oxford 5000 Series, Titanium, 50 W \\
\hline
5.2 & Graphite & Oxford 5000 Series, Titanium, 50 W \\
\hline
6.4 & Silicon (100) & Oxford 5000 Series, Iron, 50 W \\
\hline
8.0 & Germanium (111) & Oxford 5000 Series, Copper, 50 W \\
&& Head-on Hamamatsu, Copper, 2.0 W \\
\hline
9.7 & Lithium fluoride & Oxford 5000 Series, Gold, 25 W \\
\hline
Continuum & Lithium rod & Oxford 5000 Series, Tungsten, 50 W \\
&& Head-on Hamamatsu, Tungsten, 2.0 W
\end{tabular}
\caption{Energy, crystals and X-ray tubes for the generation of polarized X-rays with the ICE.} 
\label{tab:Crystals}
\end{center}
\end{table}

\begin{figure}[ht]
\begin{center}
\includegraphics[width=0.5\textwidth]{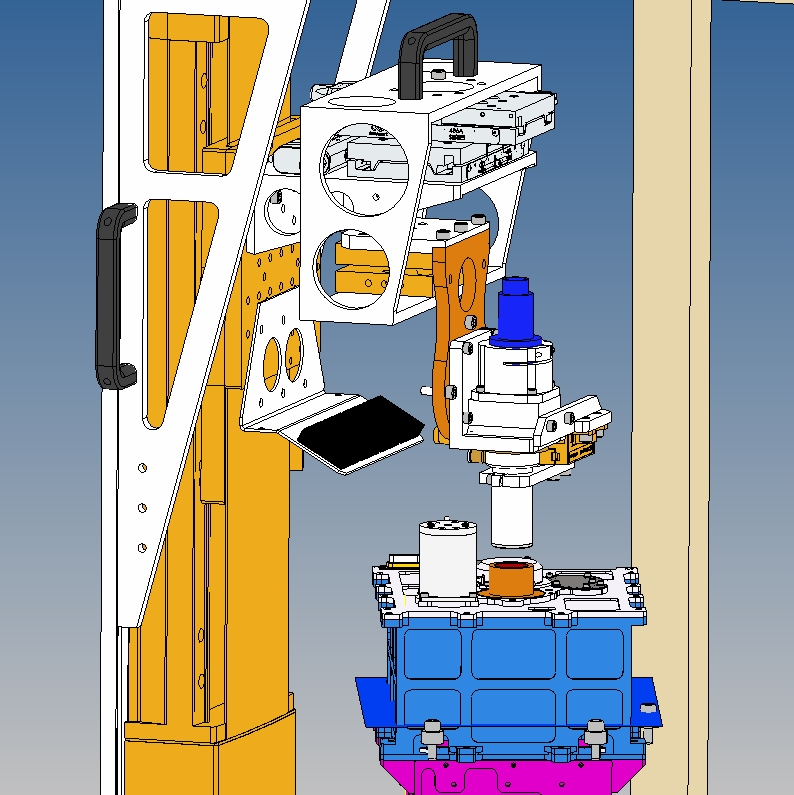}
\end{center}
\caption{Preliminary design of unpolarized calibration source based on head-on X-ray tubes.} 
\label{fig:ICE_unpol}
\end{figure}

The design of unpolarized-source setup with one of the head-on X-ray tubes is shown in 
Figure~\ref{fig:ICE_unpol}. The X-ray tube is arranged so that its direct emission illuminates the 
detector through two diaphragms with aperture between 25~$\mu$m and 2~mm. The bottom diaphragm can 
be aligned with the first and the X-ray spot of the tube with motorized stages. Helium is flown 
inside the collimator to avoid air absorption. Head-on X-ray tubes are, thanks to symmetry of the 
emission geometry, sources with a very low intrinsic polarization; for the Calcium head-on X-ray 
tube, we measured a residual polarization lower than 0.5\%. Head-on X-ray tubes with Calcium, Copper 
and Tungsten anodes are available for IXPE Instrument calibration; Calcium and Copper X-ray tubes 
have prominent fluorescence lines at 3.7 and 8.0 keV, respectively, whereas Tungsten X-ray tube does 
not have fluorescence lines between 2 and 8 keV. Many Oxford series 5000 X-ray tubes with different 
anodes will be available for Instrument calibration, but in this case radiation is emitted at about 
90 degrees and continuum emission is polarized at about 10-15\%. Continuum emission can not be 
completed removed from fluorescence lines with the spectral capabilities of the detector, and then 
Oxford 5000-Series X-ray tubes may result not appropriate for the use as unpolarized sources. If 
this will be confirmed by further tests, a backup solution will be to rotate the DU during the 
measurement to average the residual polarization.

Testing and characterization of the ICE sources will be carried out with three commercial X-ray 
detectors: (i) a CCD imager (model Andor iKon-M SY) to measure and map the beam spot and 
center the diaphragm of polarized source; (ii) a SDD spectrometer (model Amptek FAST SDD) to 
characterize the spectrum (and hence the polarization) and the counting rate of the beam. This 
detector will be also used as a reference for the efficiency measurements of the DU efficiency; 
(iii) a Si-PIN detector (model Amptek XR100CR) as a spare spectrometer.

\section{In-flight calibration} \label{sec:FCW}

Calibration sources are included in the DU to carry out in-flight calibration of the Instrument 
and monitor changes in performance, if any, during the IXPE lifetime. Variations may be due to 
changes in environmental conditions or degradation, albeit the latter are supposed to be very small 
from on-ground activities\cite{Muleri2016}. DU calibrations will be performed in orbit with the set 
of calibration sources mounted on the FCW with the aim of:
\begin{itemize}
\item monitoring the modulation factor value of the GPD for monochromatic photons and hence the 
stability of polarimetric response at two energies;
\item monitoring the energy resolution of the GPD;
\item check for the presence of spurious polarization due to, e.g., any anisotropy in the 
distribution of the background;
\item map and monitor the gain of the GEM and its non-homogeneities.
\end{itemize}

Calibration sources are hosted on a Filter and Calibration Wheel (FCW) which is included in the 
DU (see Figure~\ref{fig:FCW}). FCW has 7 different positions, which can be alternatively put in 
front of the GPD. In particular, there are:

\begin{figure}[ht]
\begin{center}
\includegraphics[width=0.8\textwidth]{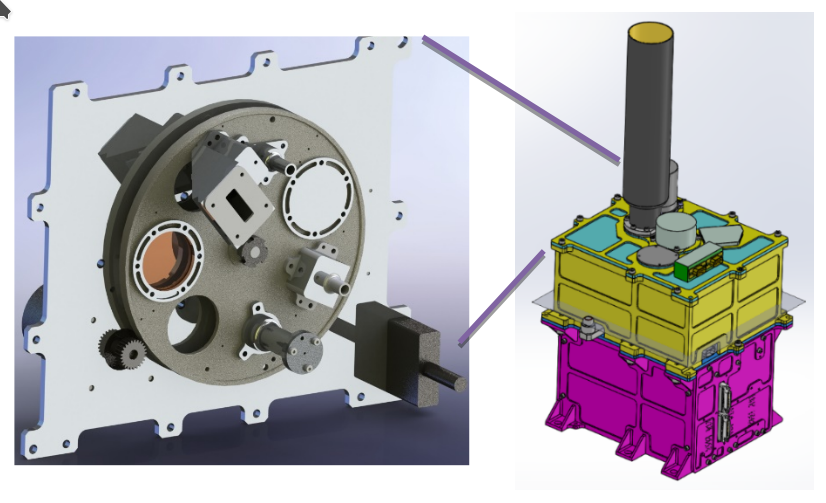}
\end{center}
\caption{Filter and Calibration Wheel in the IXPE DU.} 
\label{fig:FCW}
\end{figure}

\begin{figure}[htbp]
\begin{center}
\subfigure[]{\includegraphics[angle=0,totalheight=7cm]{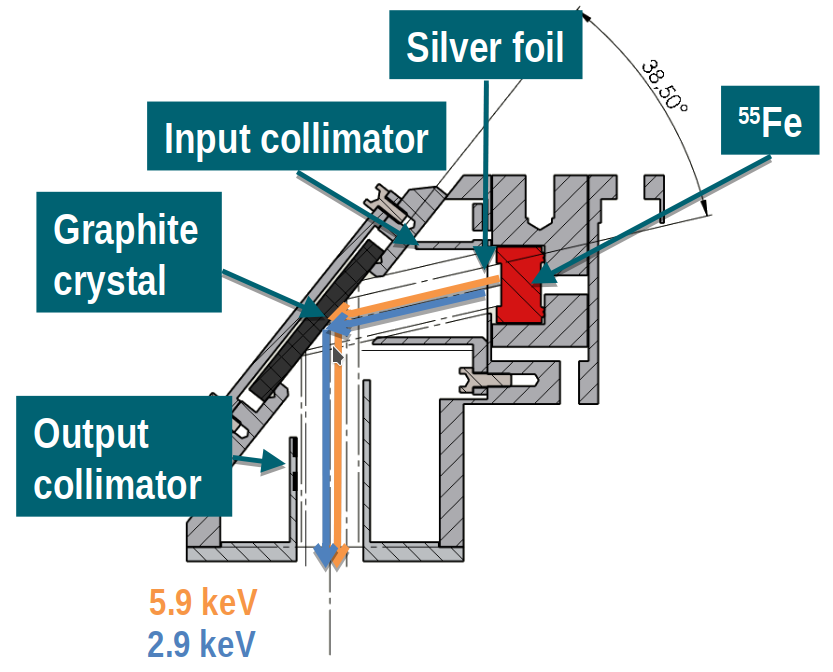}}
\hspace{3mm}
\subfigure[]{\includegraphics[angle=0,totalheight=7cm]{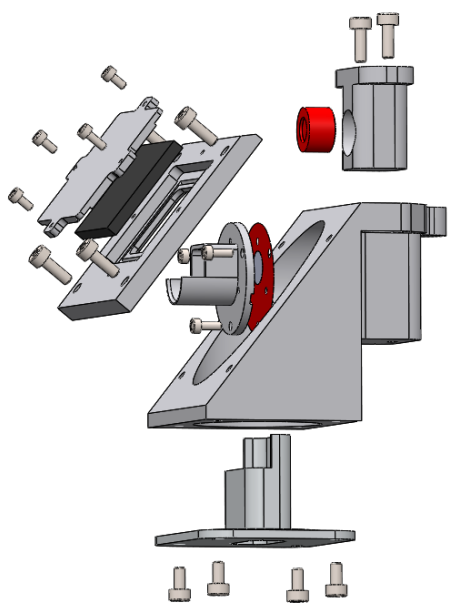}}
\end{center}
\caption{Cut-out ({\bf a}) and exploded ({\bf b}) view of Cal A.}
\label{fig:CalA}
\end{figure}

\begin{figure}[htbp]
\begin{center}
\subfigure[\label{fig:CalB}]{\includegraphics[angle=0,totalheight=5.1cm]{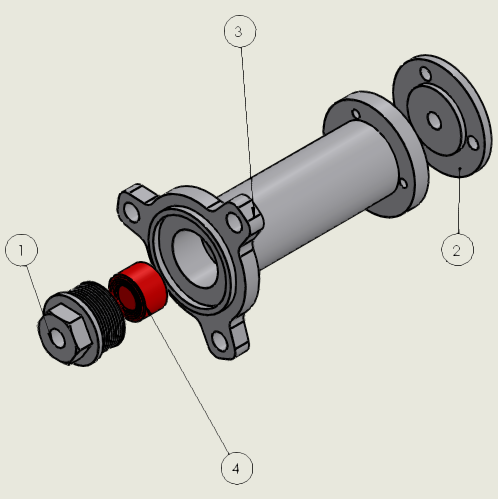}}
\hspace{3mm}
\subfigure[\label{fig:CalC}]{\includegraphics[angle=0,totalheight=5.1cm]{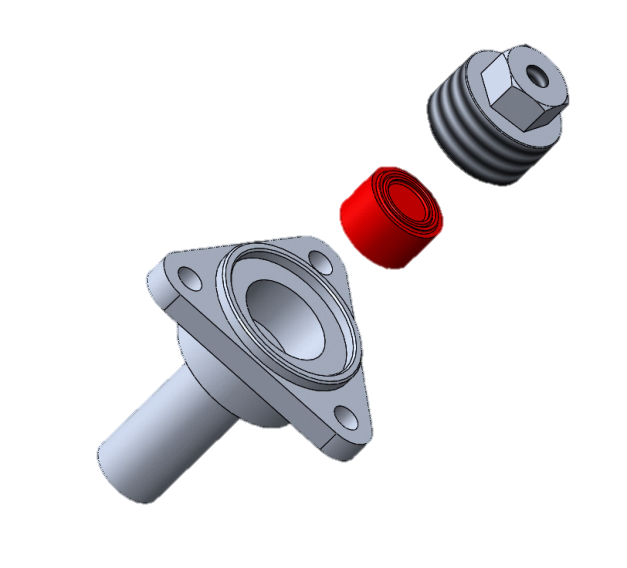}}
\hspace{3mm}
\subfigure[\label{fig:CalD}]{\includegraphics[angle=0,totalheight=5.1cm]{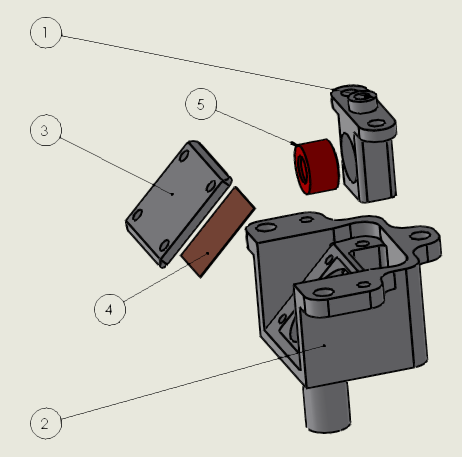}}
\end{center}
\caption{Exploded view of Cal B ({\bf a}, left panel), Cal C ({\bf b}, central panel) and Cal D 
({\bf c}, right panel).}
\end{figure}

\begin{itemize}
\item Open position. In this position no filter is put in front of the detector. The open position 
will be the standard one used for science observations. Only in case the source to observe is 
expected to be exceptionally bright (flux higher of about $4\times10^{-8}$ erg/cm$^2$/s, or 
2~Crab), the use of the gray filter described below will be recommended to reduce and better control 
the dead time of the observation. The aperture of the open position will be 31.3 mm; this value is 
derived by the requirement that not obstruction of the field of view shall occur, including the 
uncertainty in the relative positioning of the DU and of the corresponding mirror after boom 
deployment or because 
thermo-elastic deformation of the boom.
\item Close position. In this position a black filter, that is, a filter which is opaque to the 
radiation of interest, is placed in front of the detector. Mechanically, the filter will be a disk 
of aluminum 3.5~mm thick, which provides a transmission lower than 10$^{-5}$ at 12~keV.
\item Gray filter. In this position a filter partially opaque to the radiation to be observed is 
used The filter will be made of kapton 75~$\mu$m thick; in this case, the counting rate in the 
energy range 2-8 keV (or 1-12 keV) for a Crab-like spectrum (power law with index 2) will be 
reduced of a factor about 4 (or 8).
\item Calibration source A (Cal A). This source will produce polarized X-ray photons with 
precisely-known energy and polarization state, to monitor the modulation factor of the instrument 
at two energies in the IXPE energy band. A drawing of the source is shown in Figure~\ref{fig:CalA}. 
A single $^{55}$Fe nuclide is mounted and glued into a “T”-shaped holder. X-rays from $^{55}$Fe at 
5.9 keV and 6.5 keV are partially absorbed by a thin silver foil mounted in front of the $^{55}$Fe 
nuclide to produce fluorescence at 3.0~keV and 3.15~keV. Silver foil is 1.6~$\mu$m thick and it is 
deposited between two polyimide foils which are 8~$\mu$m (on the side towards the $^{55}$Fe) 
and 2~$\mu$m. Photons at 3.0 keV and 5.9 keV, collimated with a broad collimator, are 
diffracted on a graphite mosaic crystal, with FWHM mosaicity of 1.2 deg, at first and 
second order of diffraction, approximately at the same diffraction angle and hence polarization 
(see Table  4 -9 for details). A second collimator is used to block stray-light X-rays.
\item Calibration source B (Cal B). This source will produce a collimated beam of unpolarized 
photons, to monitor the absence of a spurious modulation. An exploded view of the source is 
reported in Figure~\ref{fig:CalB}. A $^{55}$Fe radioactive source (item 4) is glued in a holder 
(item 3) and screwed in a cylindrical body (item 1). At the other end of the body, a diaphragm with 
an aperture of 1 mm (item 2) collimate X-rays to produce a spot of about 3 mm on the GPD.
\item Calibration source C (Cal C). This source (see Figure~\ref{fig:CalC}) will illuminate all the 
detector sensitive area to map the gain at one energy. This source will be composed of a $^{55}$Fe 
iron radioactive source (item 2), glued in a holder (item 3) which is screwed in a body (item 1). A 
collimator allows X-ray photons to impinge on the detector sensitive area only when the source is 
in front of the GPD.
\item Calibration source D (Cal D). This source (see Figure~\ref{fig:CalD}) will illuminate all the 
detector sensitive area as Cal C, to map the gain at another energy. Cal D is based on a $^{55}$Fe 
source (item 3), glued in an aluminum holder (item 1) which illuminates a Silicon target (item 4) 
mounted on a body (item 2) to extract K$\alpha$ fluorescence from Silicon at 1.7 keV, which 
impinges on the detector. It is worth noting that the design is such that X-ray photons from 
$^{55}$Fe can not directly impinge on the GPD sensitive area to avoid detector saturation.
\end{itemize}

\begin{table}[htbp]
\begin{center}
\begin{tabular}{l|c|c}
& 3.0 keV X-rays & 5.9 keV X-rays \\
\hline
Production & Fluorescence from Ag foil & Direct emission from $^{55}$Fe \\
Diffraction angle on graphite crystal & 38.3 deg & 38.7 deg \\
Polarization of diffracted photons & 67\% & 69\% \\
Image on the detector & Strip, 4x15 mm$^2$ & Strip, 4x15 mm$^2$
\end{tabular}
\caption{Characteristics of X-rays produced by Cal A.} 
\label{tab:CalA}
\end{center}
\end{table}

All calibration sources inside the FCW contains a $^{55}$Fe nuclide, whose activity naturally 
decays with half time of 2.7 years. To minimize the activity of the on-board nuclides for 
achieving the required counting rate, all the radioactive nuclides will be inserted in the 
Instrument as late as possible in the integration flow. This will be done during Instrument 
integration on the spacecraft. Replacement of the radioactive nuclides will be possible through a 
dedicated opening in the DU lid. Radioactive sources, glued in their holders, will be extracted with 
dedicated tools and replaced with new ones in new holders. Test and dummy nuclides will be mounted 
in calibration sources for functional and environmental tests on-ground at INFN, IAPS, at MSFC and 
at Ball when the flight nuclides are not present. 

Design of calibration sources in the FCW was put to test with prototypes and counting rates with the 
activity planned for in flight nuclides are extrapolated in Table~\ref{tab:CalSources}. 

\begin{table}[htbp]
\begin{center}
\begin{tabular}{l|c|c}
Source & Activity on IXPE [mCi] & Counting rate [c/s] \\
\hline
Cal A & 100 & 5.0 at 3.0 keV \\
&& 80.0 at 5.9 keV \\
\hline
Cal B & 20 & 60 at 5.9 keV \\
\hline
Cal C & 0.5 & 100 at 5.9 keV \\
\hline
Cal D & 10 & 20 at 1.7 keV
\end{tabular}
\caption{FCW calibration sources counting rate expected at beginning of IXPE lifetime.} 
\label{tab:CalSources}
\end{center}
\end{table}

\section{Conclusions}

IXPE observatory will open a new window in X-ray astronomy and therefore its calibration will have 
to rely mainly on the characterization with laboratory sources. Calibration of the IXPE payload 
will be modular: mirror and focal plane instrument will be first calibrated separately and then 
jointly. On-ground Instrument calibration will be performed at INAF-IAPS in Italy and it will be 
dedicated to an extensive characterization of the GPD. Instrument calibration equipment will include 
polarized and unpolarized sources, whose custom design has been refined for 10 years during the 
development of the GPD. In-flight calibration will be possible thanks to four calibration sources 
which will be placed in front of the GPD when required by rotating a Filter and Calibration Wheel. 
Monitoring of the modulation factor at 3.0 and 5.9 keV will be possible with a single source of 
polarized X-rays. Absence of systematic effect will be verified at 5.9 keV, whereas gain monitoring 
and mapping will be possible at two energies, 1.7 and 5.9 keV.

\acknowledgments

Italian contribution to IXPE mission is supported by the Italian Space Agency through agreement 
ASI-INAF n.2017-12-H.0 and ASI-INFN agreement n.2017-13-H.0.

\bibliography{References}

\begin{thebibliography}{1}

\bibitem{Odell2018}
{O'Dell}, S. o. b. o. t. I.~C., ``{The Imaging X-ray Polarimetry Explorer
  (IXPE): Overview},'' in [{\em Proc. of SPIE}{\nolinebreak\hspace{0.1em}]},
  {\bf 10699} (2018).

\bibitem{Costa2001}
{Costa}, E., {Soffitta}, P., {Bellazzini}, R., {Brez}, A., {Lumb}, N., and
  {Spandre}, G., ``{An efficient photoelectric X-ray polarimeter for the study
  of black holes and neutron stars},'' {\em \nat}~{\bf 411},  662 (2001).

\bibitem{Bellazzini2006}
{Bellazzini}, R., {Spandre}, G., {Minuti}, M., {Baldini}, L., {Brez}, A.,
  {Cavalca}, F., {Latronico}, L., {Omodei}, N., {Massai}, M.~M., {Sgro'}, C.,
  {Costa}, E., {Soffitta}, P., {Krummenacher}, F., and {de Oliveira}, R.,
  ``{Direct reading of charge multipliers with a self-triggering CMOS analog
  chip with 105 k pixels at 50 {$\mu$}m pitch},'' {\em Nuclear Instruments and
  Methods in Physics Research A}~{\bf 566},  552 (2006).

\bibitem{Bellazzini2007}
{Bellazzini}, R., {Spandre}, G., {Minuti}, M., {Baldini}, L., {Brez}, A.,
  {Latronico}, L., {Omodei}, N., {Razzano}, M., {Massai}, M.~M.,
  {Pesce-Rollins}, M., {Sgr{\'o}}, C., {Costa}, E., {Soffitta}, P., {Sipila},
  H., and {Lempinen}, E., ``{A sealed Gas Pixel Detector for X-ray
  astronomy},'' {\em Nuclear Instruments and Methods in Physics Research
  A}~{\bf 579},  853 (2007).

\bibitem{Soffitta2013}
{Soffitta}, P., {Barcons}, X., {Bellazzini}, R., {Braga}, J., {Costa}, E.,
  {Fraser}, G.~W., {Gburek}, S., {Huovelin}, J., {Matt}, G., {Pearce}, M.,
  {Poutanen}, J., {Reglero}, V., {Santangelo}, A., {Sunyaev}, R.~A.,
  {Tagliaferri}, G., {Weisskopf}, M., {Aloisio}, R., {Amato}, E., {Attin{\'a}},
  P., {Axelsson}, M., {Baldini}, L., {Basso}, S., {Bianchi}, S., {Blasi}, P.,
  {Bregeon}, J., {Brez}, A., {Bucciantini}, N., {Burderi}, L., {Burwitz}, V.,
  {Casella}, P., {Churazov}, E., {Civitani}, M., {Covino}, S., {Curado da
  Silva}, R.~M., {Cusumano}, G., {Dadina}, M., {D'Amico}, F., {De Rosa}, A.,
  {Di Cosimo}, S., {Di Persio}, G., {Di Salvo}, T., {Dovciak}, M., {Elsner},
  R., {Eyles}, C.~J., {Fabian}, A.~C., {Fabiani}, S., {Feng}, H., {Giarrusso},
  S., {Goosmann}, R.~W., {Grandi}, P., {Grosso}, N., {Israel}, G., {Jackson},
  M., {Kaaret}, P., {Karas}, V., {Kuss}, M., {Lai}, D., {Rosa}, G.~L.,
  {Larsson}, J., {Larsson}, S., {Latronico}, L., {Maggio}, A., {Maia}, J.,
  {Marin}, F., {Massai}, M.~M., {Mineo}, T., {Minuti}, M., {Moretti}, E.,
  {Muleri}, F., {O'Dell}, S.~L., {Pareschi}, G., {Peres}, G., {Pesce}, M.,
  {Petrucci}, P.-O., {Pinchera}, M., {Porquet}, D., {Ramsey}, B., {Rea}, N.,
  {Reale}, F., {Rodrigo}, J.~M., {R{\'o}{\.z}a{\'n}ska}, A., {Rubini}, A.,
  {Rudawy}, P., {Ryde}, F., {Salvati}, M., {de Santiago}, V.~A., {Sazonov}, S.,
  {Sgr{\'o}}, C., {Silver}, E., {Spandre}, G., {Spiga}, D., {Stella}, L.,
  {Tamagawa}, T., {Tamborra}, F., {Tavecchio}, F., {Teixeira Dias}, T., {van
  Adelsberg}, M., {Wu}, K., and {Zane}, S., ``{XIPE: the X-ray imaging
  polarimetry explorer},'' {\em Experimental Astronomy}~{\bf 36},  523 (2013).

\bibitem{Tamagawa2009}
{Tamagawa}, T., {Hayato}, A., {Asami}, F., {Abe}, K., {Iwamoto}, S.,
  {Nakamura}, S., {Harayama}, A., {Iwahashi}, T., {Konami}, S., {Hamagaki}, H.,
  {Yamaguchi}, Y.~L., {Tawara}, H., and {Makishima}, K., ``{Development of
  thick-foil and fine-pitch GEMs with a laser etching technique},'' {\em
  Nuclear Instruments and Methods in Physics Research A}~{\bf 608},  390
  (2009).

\bibitem{Muleri2008b}
{Muleri}, F., {Soffitta}, P., {Bellazzini}, R., {Brez}, A., {Costa}, E.,
  {Frutti}, M., {Mastropietro}, M., {Morelli}, E., {Pinchera}, M., {Rubini},
  A., and {Spandre}, G., ``{A versatile facility for the calibration of X-ray
  polarimeters with polarized and unpolarized controlled beams},'' in [{\em
  Proc. of SPIE}{\nolinebreak\hspace{0.1em}]},   {\bf 7011},  701127 (2008).

\bibitem{Muleri2016}
{Muleri}, F., {Soffitta}, P., {Baldini}, L., {Bellazzini}, R., {Brez}, A.,
  {Costa}, E., {Di Lalla}, N., {Del Monte}, E., {Evangelista}, Y., {Latronico},
  L., {Manfreda}, A., {Minuti}, M., {Pesce-Rollins}, M., {Pinchera}, M.,
  {Rubini}, A., {Sgr{\`o}}, C., {Spada}, F., and {Spandre}, G., ``{Performance
  of the Gas Pixel Detector: an x-ray imaging polarimeter for upcoming missions
  of astrophysics},'' in [{\em Space Telescopes and Instrumentation 2016:
  Ultraviolet to Gamma Ray}{\nolinebreak\hspace{0.1em}]},  {\em \procspie} {\bf
  9905},  99054G (2016).

\end{thebibliography}
\bibliographystyle{spiebib}

\end{document}